\documentclass[english,iop]{emulateapj}
\usepackage[T1]{fontenc}
\usepackage{amssymb}
\usepackage{graphicx}

\makeatletter

\DeclareRobustCommand{\greektext}{%
  \fontencoding{LGR}\selectfont\def\encodingdefault{LGR}}
\DeclareRobustCommand{\textgreek}[1]{\leavevmode{\greektext #1}}
\DeclareFontEncoding{LGR}{}{}
\DeclareTextSymbol{\~}{LGR}{126}

\usepackage{apjfonts}
\usepackage[hyperfootnotes=false]{hyperref}
\hypersetup{colorlinks=true,citecolor=blue}
\shorttitle{Comparison of the H$\alpha$ and FUV continuum backgrounds}
\shortauthors{SEON ET AL.}

\makeatother

\usepackage{babel}
\begin{document}

\title{Comparison of the diffuse H$\alpha$ and FUV continuum backgrounds:\\
On the Origins of the diffuse H$\alpha$ background}

\author{Kwang-Il Seon\altaffilmark{1}, Adolf Witt\altaffilmark{2}, Il-Joong
Kim\altaffilmark{1}, Jong-Ho Shinn\altaffilmark{1}, \\Jerry Edelstein\altaffilmark{3},
Kyoung-Wook Min\altaffilmark{4}, and Wonyong Han\altaffilmark{1}}

\altaffiltext{1}{Korea Astronomy and Space Science Institute, Daejeon 305-348, Republic of Korea; kiseon@kasi.re.kr}
\altaffiltext{2}{Ritter Astrophysical Research Center, University of Toledo, Toledo, OH, USA}
\altaffiltext{3}{Space Sciences Laboratory, University of California, Berkeley, CA 94702, USA}
\altaffiltext{4}{Korea Advanced Institute of Science and Technology, Daejeon 305-701, Republic of Korea} 
\begin{abstract}
We compare the diffuse H$\alpha$ map of our Galaxy with the FUV (1370--1710\AA)
continuum map. The H$\alpha$ intensity correlates well with the FUV
intensity. The H$\alpha$/FUV intensity ratio increases in general
with the H$\alpha$ intensity and the FUV hardness ratio (1370--1520\AA\
to 1560--1710\AA), implying that late OB stars may be the main source
of the H$\alpha$ recombination line at high latitudes. The variation
of the H$\alpha$ intensity as a function of the Galactic latitude
is also very similar to that of the FUV intensity. The results likely
suggest that not only the original radiation sources of the H$\alpha$
and FUV backgrounds but also the radiative transfer mechanisms responsible
for the diffuse backgrounds are largely common. Therefore, we propose
a scenario wherein the H$\alpha$ background at high latitudes is
mostly composed of two components, H$\alpha$ photons produced by
in-situ recombination at the ionized regions around late OB stars
and dust-scattered light of the H$\alpha$ photons originating from
late OB stars.
\end{abstract}


\section{Introduction}

The diffuse H$\alpha$ background is generally believed to originate
from the extended, nearly fully ionized regions (called warm ionized
medium, or WIM) along the lines of sight (\citealt{Reynolds98}; see
\citealt{Haffner2009} for a recent review). The volume filling fraction
of the WIM increases from $\sim0.1$ at the midplane to $>0.3-0.4$
at $|z|=1$ kpc \citep{Kulkarni1987,Reynolds1991,Peterson2002}. The
large energy requirement of the WIM strongly suggests that O stars
in the Galactic plane are the primary ionization source \citep{Reynolds84,Reynolds90}.
There have been numerous investigations about how the Lyman continuum
(Lyc) photons can escape the immediate vicinity of the O stars (or
the bright \ion{H}{2} regions) in the Galactic plane and ionize the
diffuse interstellar medium (ISM) \citep{MillerCox_1993,DoveShull_1994,Seon2009a,Wood2010}.
However, the question of how the Lyc photons can propagate almost
freely up to the scaleheight ($\sim1$ kpc) of the WIM has not been
fully resolved because the photoionization models require the presence
of very low density paths through the ISM.

In this paper, we explore two alternatives to explain the diffuse
H$\alpha$ emission of our Galaxy. The first scenario, which was investigated
for external galaxies, is that late, field OB stars outside the bright
\ion{H}{2} regions ionize some of the WIM, reducing the required
Lyc leakage from the bright \ion{H}{2} regions in the galactic plane
\citep{Hoopes00,Hoopes01}. The mean spectral type of stars in the
field is later than that in \ion{H}{2} regions \citep{Massey1995,Hoopes00}.
As the H$\alpha$ intensity is directly proportional to Lyc intensity
in the ionization bounded regions and the later-type stars would produce
a lower ratio of ionizing Lyc to non-ionizing far-ultraviolet (FUV)
luminosities, the H$\alpha$/FUV intensity ratio can constrain the
spectral type of the ionizing stars. \citet{Hoopes00} and \citet{Hoopes01}
measured H$\alpha$/FUV intensity ratios for 10 nearby spiral galaxies,
and concluded that late OB stars in the field are indeed an important
source of ionization in those galaxies.

Secondly, we note that the majority of the H$\alpha$ excess intensity
in a number of high-latitude clouds in our Galaxy can be understood
as light scattered off the interstellar dust \citep{Mattila2007,Lehtinen2010,Witt2010}.
The possible presence of scattered H$\alpha$ radiation at high-latitude
clouds was first considered by \citet{Jura1979}. Photoionization
by an external field of Lyc is expected to result in ionization of
a thin outer shell of an isolated interstellar cloud while the observations
of the H\textgreek{a} light from the high-latitude clouds clearly
showed the dense inner core of the cloud as the brightest feature.
\citet{Witt2010} argued that both a substantial fraction of the diffuse
high-latitude H$\alpha$ background intensity and much of the variance
in the high-latitude H$\alpha$ background are due to scattering by
interstellar dust of H$\alpha$ photons originating elsewhere in the
Galaxy. If this is the case, the diffuse H$\alpha$ emission should
be closely related to the far-ultraviolet (FUV) continuum background,
which is mostly dust-scattered light of the FUV stellar radiation
\citep{Bowyer91,Seon2011}. 

Therefore, comparison of the diffuse H$\alpha$ emission with the
FUV continuum background could not only provide a clue to the spectral
type of ionizing sources of the WIM in our Galaxy but also confirm
the significance of dust-scattering in the diffuse H$\alpha$. In
fact, \citet{Seon2011} found a strong correlation between the diffuse
FUV (1370--1710\AA) and H$\alpha$ backgrounds, suggesting a similarity
between the origins of both backgrounds. In this paper, we present
a more detailed comparison of the Galactic H$\alpha$ map with the
FUV continuum map and propose a scenario on the origin of the diffuse
H$\alpha$ emission. We describe the data used in the present study
in Section 2. In Section 3, the spectral type of the stars responsible
for the diffuse H$\alpha$ emission is investigated. Section 4 contains
a discussion on the dust-scattering of H$\alpha$ photons. We discuss
the main result in Section 5. A brief summary is presented in Section
6.

\section{Data}

The FUV data was obtained by the \emph{Spectroscopy of Plasma Evolution
from Astrophysical} \emph{Radiation (SPEAR)} instruments, also known
as \emph{Far Ultraviolet Imaging Spectrograph (FIMS}) \citep{Edelstein06a,Edelstein06b,Seon2011}.
The \emph{SPEAR/FIMS} instrument is a dual-channel FUV imaging spectrograph
(``Short'' wavelength channel {[}S-band{]} 900-1150\AA, ``Long''
wavelength channel {[}L-band{]} 1350--1750\AA; $\lambda/\Delta\lambda\sim550$)
with a large field of view (S-band, $4^{\circ}\times4'.6$; L-band,
$7^{\circ}.4\times4'.3$) and $10'$ imaging resolution onboard the
first Korean astronomical satellite \emph{STSAT-1}, designed to observe
diffuse FUV emission. The \emph{SPEAR/FIMS} survey observations were
performed by scanning the sky at constant ecliptic longitude from
the north ecliptic pole to the south ecliptic pole, during the eclipses.
The pole-to-pole scanning drifted 360$^{\circ}$ along the ecliptic
equator for one year because of the properties of a sun-synchronous
orbit. The mission observed $\sim80$\% of the sky.

The diffuse FUV map in L-band was obtained by eliminating stars from
the total FUV sky map and by binning photon events using the \emph{Hierarchical
Equal Area isoLatitude Pixelization (HEALPix)} tessellation scheme
\citep{Gorski05} with $\sim1^{\circ}$ pixels (resolution parameter
$N_{{\rm side}}=64$). We excluded the wavelength regions of the strongest
emission lines (\ion{Si}{4} $\lambda$1398, \ion{Si}{2}{*} $\lambda$1532,
\ion{C}{4} $\lambda\lambda$1548, 1551, and \ion{Al}{2} $\lambda$1671)
and averaged the data in the wavelength regions of 1370--1520\AA,
1560--1660\AA, and 1680--1710\AA. A detailed description of the
FUV continuum map is presented in \citet{Seon2011}.

The all-sky survey maps in H$\alpha$ and 100 $\mu$m were obtained
from \citet{Finkbeiner03} and \citet{Schlegel1998}, respectively.
The all-sky map of \ion{H}{1} column density is from the Leiden/Argentine/Bonn
(LAB) Survey \citep{Kalberla2005}. The data interpolated onto a \emph{HEALPix}
projection were obtained from the Legacy Archive for Microwave Background
Data Analysis (LAMBDA).

\section{Spectral Type of Ionizing Sources}

\begin{figure}[t]
\begin{centering}
\includegraphics[scale=0.6,angle=90]{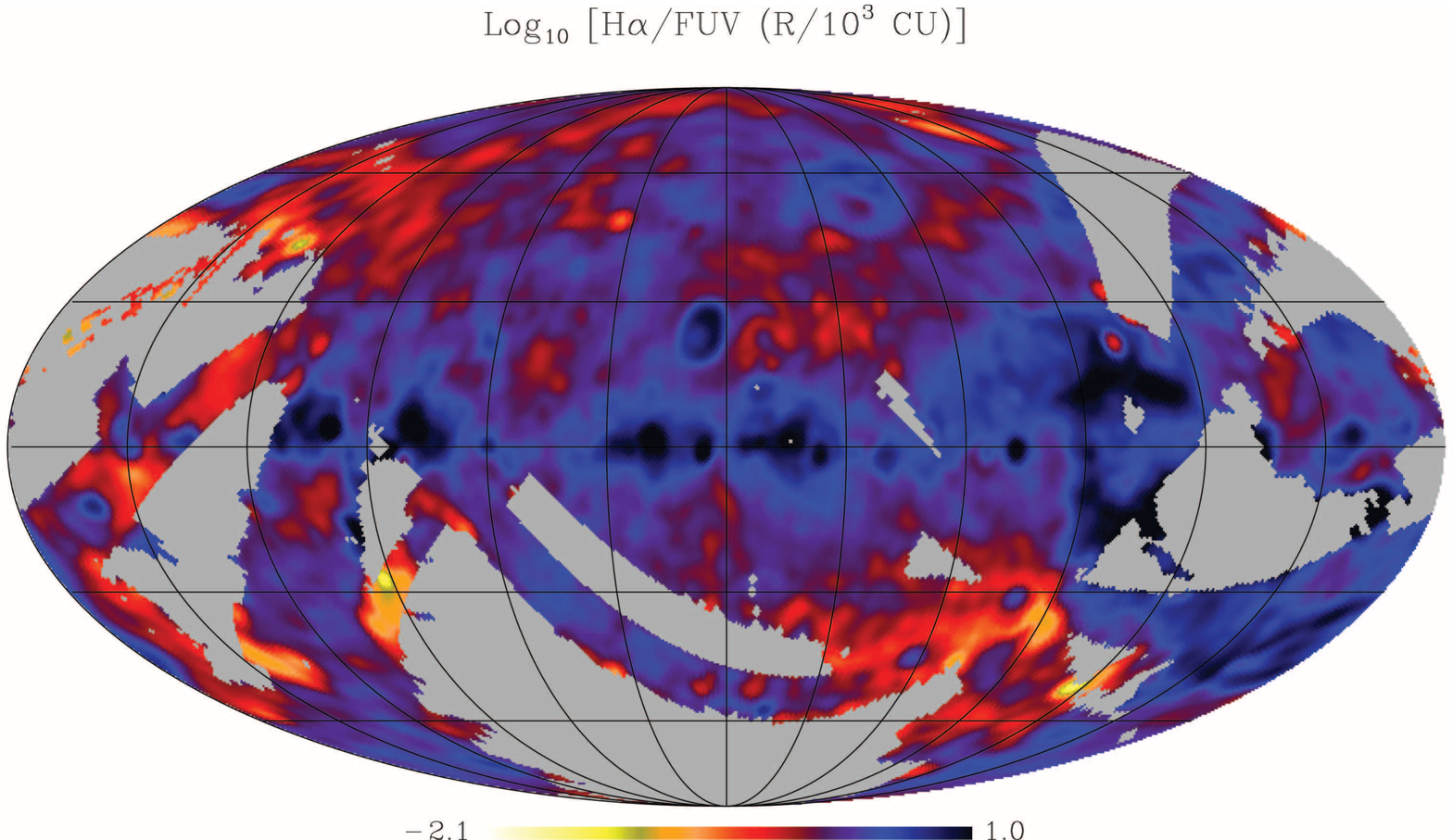}\bigskip{}

\par\end{centering}

\begin{centering}
\bigskip{}

\par\end{centering}

\begin{centering}
\includegraphics[scale=0.6,angle=90]{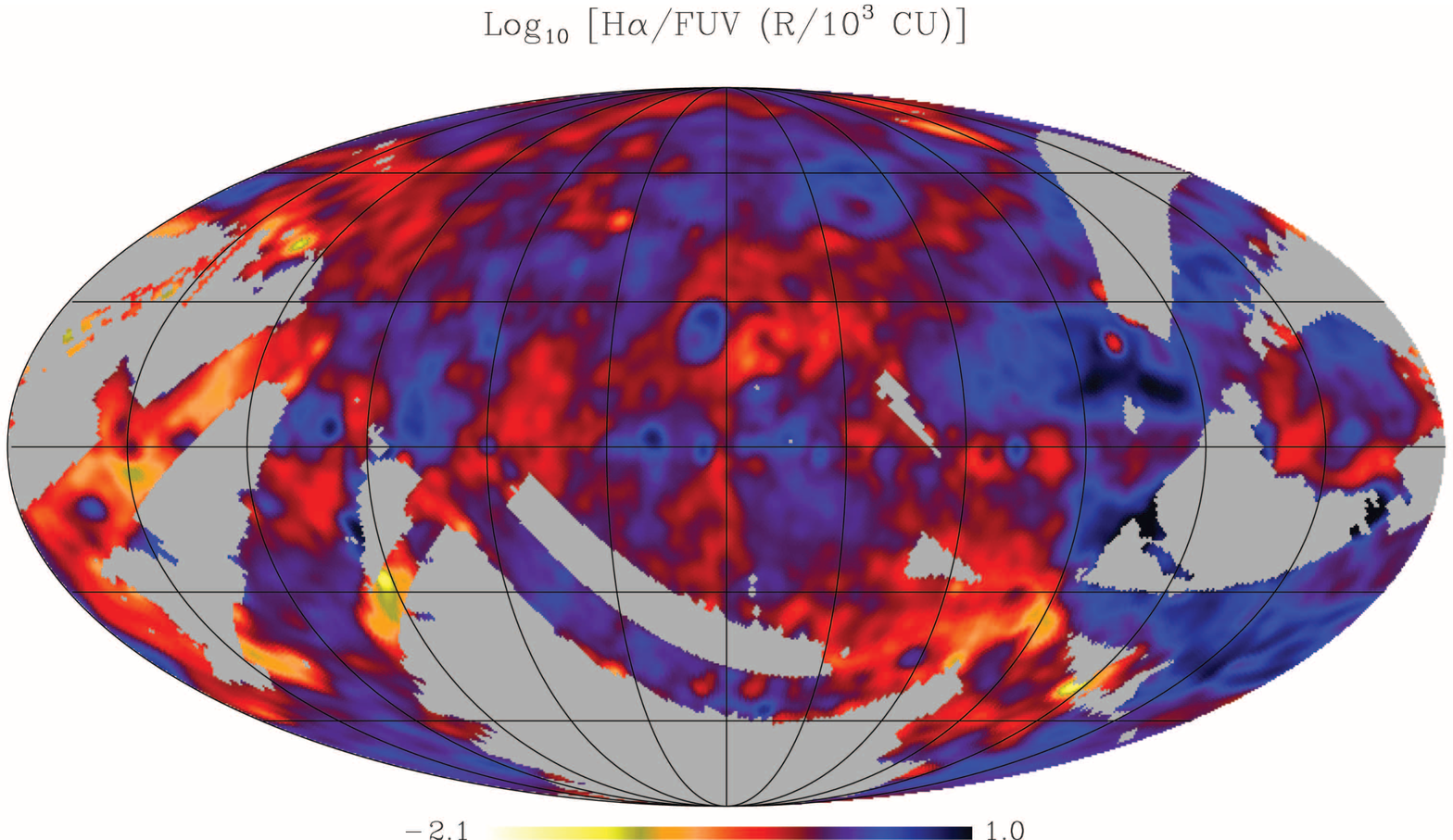}
\par\end{centering}

\caption{\label{map_ratio}Mollweide projections of the diffuse H$\alpha$
to FUV intensity ratio maps. Top panel shows the ratio map before
dust-extinction correction. Bottom panel shows the ratio map after
dust-extinction correction. The dust-extinction correction was performed
assuming the uniformly mixed gas and dust. Here, 1 R (rayleigh) =
$10^{6}/4\pi$ photons cm$^{-2}$ s$^{-1}$ sr$^{-1}$ and 1 CU (continuum
unit) = photons cm$^{-2}$ s$^{-1}$ sr$^{-1}$ \AA$^{-1}$. Galactic
coordinates centered at $(l,b)=(0^{\circ},0^{\circ})$ with longitude
increasing toward the left are shown with latitude and longitude lines
on a 30$^{\circ}$ grid.}
\end{figure}

Figure \ref{map_ratio} shows the H$\alpha$/FUV intensity ratio maps
obtained by dividing the H$\alpha$ intensity with the FUV continuum
intensity and smoothed with a spherical Gaussian function with a full-width
at half maximum (FWHM) $\sim3^{\circ}$ \citep{Seon2006}. The smoothing
was performed using the spherical harmonic transform in the \emph{HEALPix}
scheme so to smear the same surface areas in all directions. We excluded
the regions with low signal-to-noise ratio (S/N $<3$) and two regions
between edges of observing strips centered at $(l,b)\sim(45^{\circ},-35^{\circ})$
and $\sim(225^{\circ},45^{\circ})$, which appeared to produce an
artifact with very high H$\alpha$/FUV intensity ratios after the
smoothing. The FUV and H$\alpha$ intensities are expressed in units
of rayleighs (1 rayleigh {[}R{]} $=10^{6}/4\pi$ photons cm$^{-2}$
s$^{-1}$ sr$^{-1}$) and continuum unit (1 CU = photons cm$^{-2}$
s$^{-1}$ sr$^{-1}$ \AA$^{-1}$). In the figure, the region with
higher H$\alpha$/FUV ratio is represented in blue. Top and bottom
panels show the intensity ratio maps before and after dust-extinction
correction, respectively. Using the standard approach of \citet{Bennett2003}
for the emitting medium coexisting with dust and the extinction map
of \citet{Schlegel1998}, the correction for dust-extinction was performed
for both radiations. In the top panel of the figure, it is clear that
bright \ion{H}{2} regions have higher H$\alpha$/FUV ratios than
the diffuse H$\alpha$ regions. Correction for dust-extinction lowers
the ratio, because the correction for FUV is larger than for H\textgreek{a}.
As dust-extinction is stronger at low latitudes, the correction for
H II regions is bigger than for the WIM. Therefore, the distinction
between the bright \ion{H}{2} and the WIM regions becomes less clear
after the dust-extinction correction, as shown in the bottom panel.

For the quantitative comparison, the WIM regions (we will use the
term WIM to denote the diffuse H$\alpha$ regions that are not related
to the bright \ion{H}{2} regions, regardless of physical origin)
were isolated in the H$\alpha$ image using a masking procedure similar
to that described in \citet{Hoopes01}. The H$\alpha$ image was smoothed
with a median filter using a pixel size of $\sim15^{\circ}$ which
corresponds approximately to the size of the largest \ion{H}{2} regions.
The smoothed image was subtracted from the original, leaving an image
of the small-scale structures only. In other words, the smoothly varying
component was subtracted. On this image a mask was created by assigning
the value `1' to the pixels with intensity lower than a prescribed
threshold and `0' to the others. We tried several threshold values
between 0.1 and 50 R as the separation between the WIM and \ion{H}{2}
regions was rather fuzzy. The unsmoothed H$\alpha$ image was then
multiplied by the mask, leaving an image of only the WIM. A WIM FUV
image was also created using the same mask. The H$\alpha$ and FUV
images for \ion{H}{2} regions were created by multiplying the inverted
WIM mask.

\begin{figure}[t]
\begin{centering}
\includegraphics[scale=0.6]{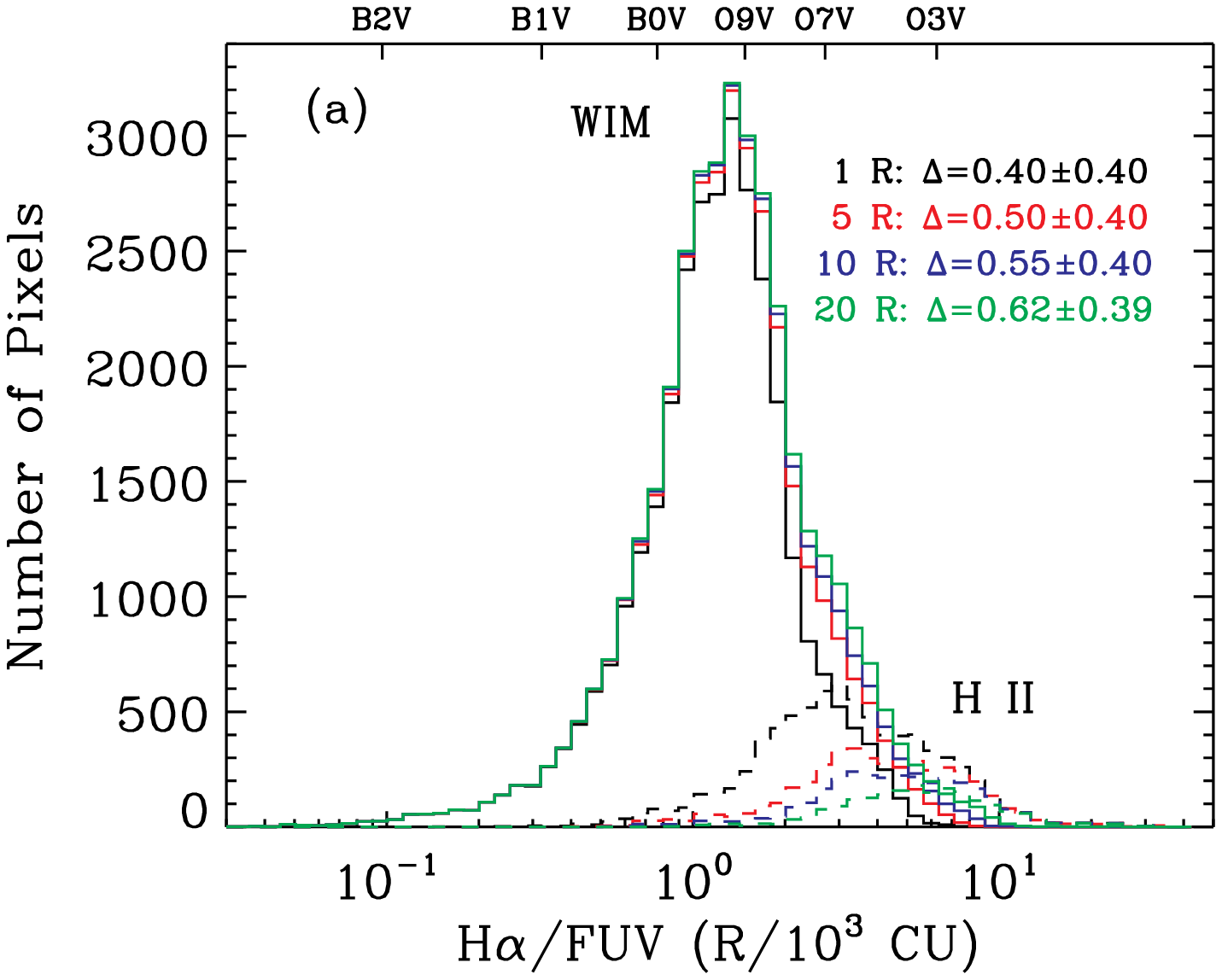}\hfill{}\includegraphics[scale=0.6]{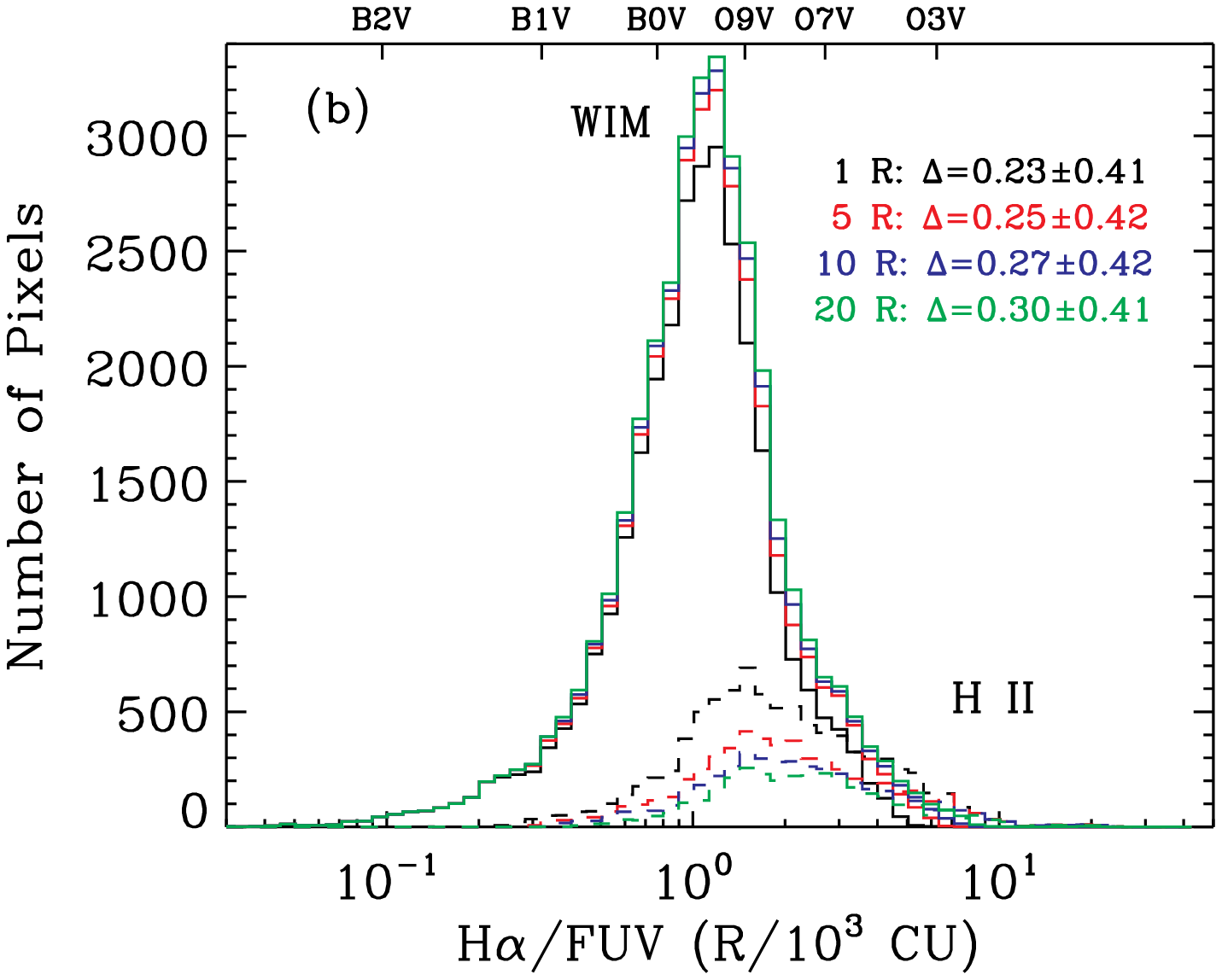}
\par\end{centering}

\caption{\label{hist_ratio}Histograms of the H$\alpha$/FUV ratios for the
\ion{H}{2} regions and WIM. Extinction was (a) not corrected and
(b) corrected assuming that the emitting gas is coextensive with dust.
$\Delta$ indicates the difference of the mean intensity ratios in
logarithmic scale for various thresholds. The ratios calculated for
various stellar types are also indicated.}
\end{figure}

Figure \ref{hist_ratio}(a) shows the histograms of the H$\alpha$/FUV
intensity ratios for the WIM and \ion{H}{2} regions in our Galaxy,
with the thresholds of 1, 5, 10 and 20 R. The threshold `1 R' masks
most of the \ion{H}{2} regions and `20 R' the brightest ones. The
mean ratios for the \ion{H}{2} regions are generally higher than
the mean ratios in the WIM. The difference in the logarithm of the
average ratio between the \ion{H}{2} regions and the WIM increases
smoothly from $0.40\pm0.40$ to $0.62\pm0.39$ dex as the threshold
increases from 1 to 20 R. These values are smaller than those found
in 10 external galaxies in \citet{Hoopes01}, although the largest
value of $\sim0.62$ is consistent with the results for NGC 2903,
NGC 5457, and NGC 1512 within the error ranges. Figure \ref{hist_ratio}(b)
shows the histograms of the H$\alpha$/FUV intensity ratios after
dust-extinction correction. Correction for dust-extinction lowers
the difference in the average ratio, as in the bottom panel of Figure
\ref{map_ratio}.

The smaller difference of the mean ratios in our Galaxy than those
of other galaxies may be attributable to geometrical effects, as the
WIM and \ion{H}{2} regions overlap in our Galaxy while in the face-on
external galaxies examined in \citet{Hoopes00} and \citet{Hoopes01}
two regions are better separated from each other. The field stars
in the Galactic plane overlapping with \ion{H}{2} regions would lower
the H$\alpha$/FUV intensity ratio observed toward the bright \ion{H}{2}
regions. Some of the H$\alpha$ and FUV photons originating from \ion{H}{2}
regions are scattered into high latitudes \citep{Wood1999,Witt2010,Seon2011}.
The dust-scattering of H$\alpha$ and FUV photons from the \ion{H}{2}
regions to high latitudes would then increase the ratio for the WIM,
further lowering the difference of the ratios. Furthermore, all the
\ion{H}{2} regions are at the same distance in a face-on external
galaxy while in the Milky Way the apparent size of \ion{H}{2} regions
is potentially dominated by their distances making the separation
more difficult. Therefore, the difference between the \ion{H}{2}
regions and the WIM in our Galaxy may not be significantly different
from other galaxies when the Galaxy is viewed face-on.

Along the top axis of Figure \ref{hist_ratio} are shown the stellar
spectral types that would yield the corresponding H$\alpha$/FUV ratios.
We derived a temperature and effective gravity for given spectral
types using the calibration of \citet{Vacca1996}. We then calculated
the Lyc and FUV photon luminosities $L_{{\rm Lyc}}$ and $L_{{\rm FUV}}$
for spectral types of B2--O3 stars in the wavebands $\lambda<912$\AA\
and $1370-1710$\AA, respectively, by interpolating on a grid of
Kurucz models \citep{Castelli03}. The H$\alpha$ luminosities $L_{{\rm H\alpha}}$
equal $0.46L_{{\rm Lyc}}$, assuming case B recombination. Since \citet{Vacca1996}
covers spectral types only earlier than or equal to B0.5, we extrapolated
the calibration down to B2. The temperatures estimated in this way
were $\sim5-19$\% higher than the values obtained from the older
calibration of \citet{Straizys81}. The extraporation to B2 would
be accurate within $\sim19$\%. Figure \ref{hist_ratio} suggests
that the difference between the H$\alpha$/FUV ratios in the WIM and
those in the \ion{H}{2} regions is due to the difference in main
ionizing sources.

An interesting result in Figure \ref{hist_ratio} is that the median
value of the observed ratios in the WIM corresponds to a value predictable
from $\sim$ O9 -- B0 stars with effective temperatures of $\sim$
35,000 K. \citet{Sembach2000} modeled the WIM as a combination of
overlapping \ion{H}{2} regions and found that typical values of the
line ratios [\ion{S}{2}] $\lambda$6716/H$\alpha$ $\sim0.2-0.4$,
[\ion{N}{2}] $\lambda$6583/H$\alpha$ $\sim0.3-0.6$, [\ion{O}{1}]
$\lambda$6300/H$\alpha$ $\sim0.01-0.03$, [\ion{O}{3}] $\lambda$5007/H$\alpha$
$\lesssim0.1$, and \ion{He}{1} $\lambda$5876/H$\alpha$ $<0.02$
observed in the Galactic WIM can be well reproduced by an ionizing
spectrum of a star with $T_{{\rm eff}}\sim35,000$ K; \citet{Domgorgen94}
also found that the WIM observations best agree with $T_{{\rm eff}}\sim38,000$
K. The results are consistent with the H$\alpha$/FUV ratios, thereby
strongly supporting that the H$\alpha$ emission originates from late
OB-type stars. The optical line ratios will be discussed more in Section
5. We also note that the H$\alpha$/FUV ratio has a well-defined limiting
value corresponding to B2 stars, meaning that the latest stellar type
that is capable of ionizing the WIM may be about B2.

\begin{figure}[t]
\begin{centering}
\includegraphics[scale=0.5]{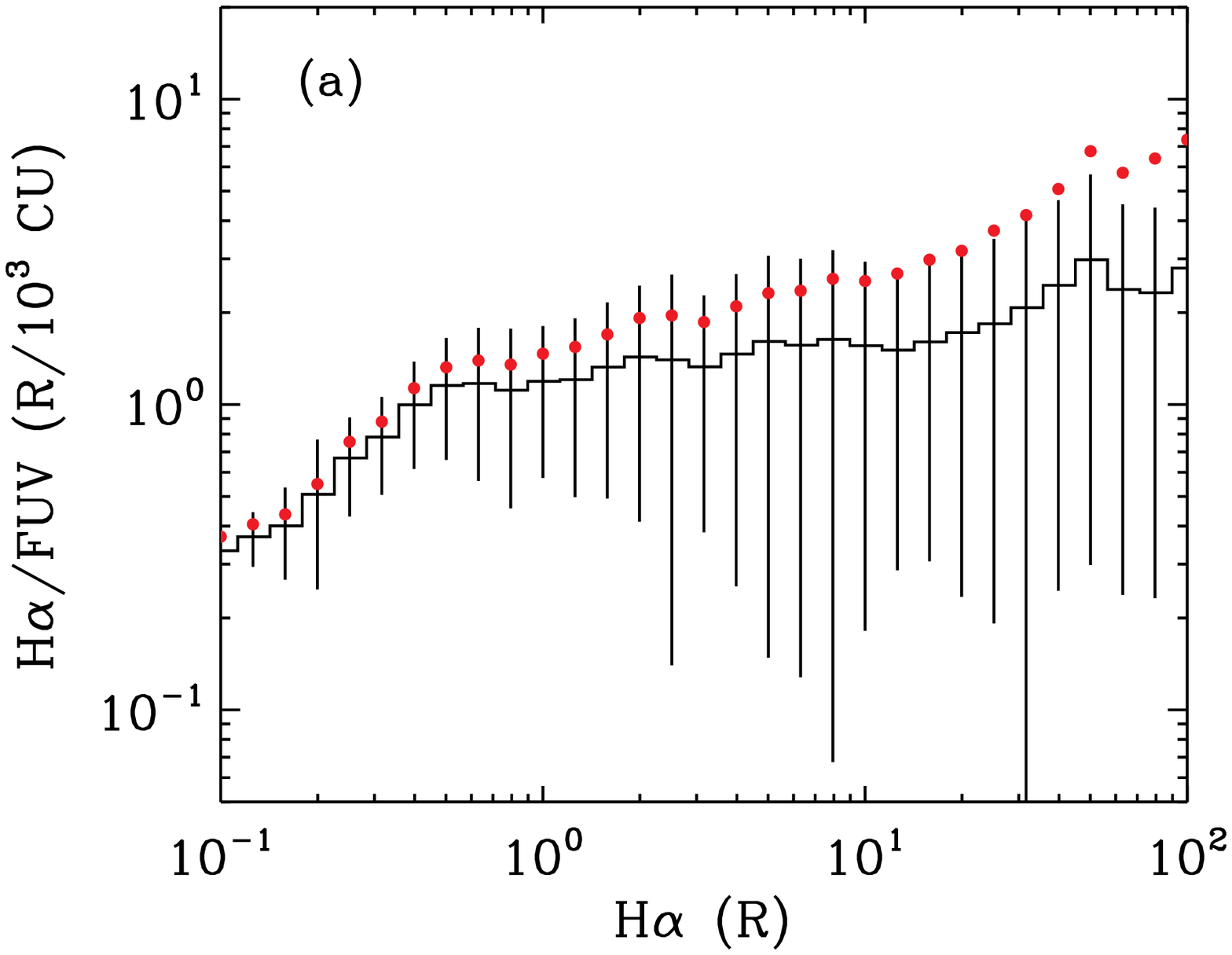}\hfill{}\includegraphics[clip,scale=0.5]{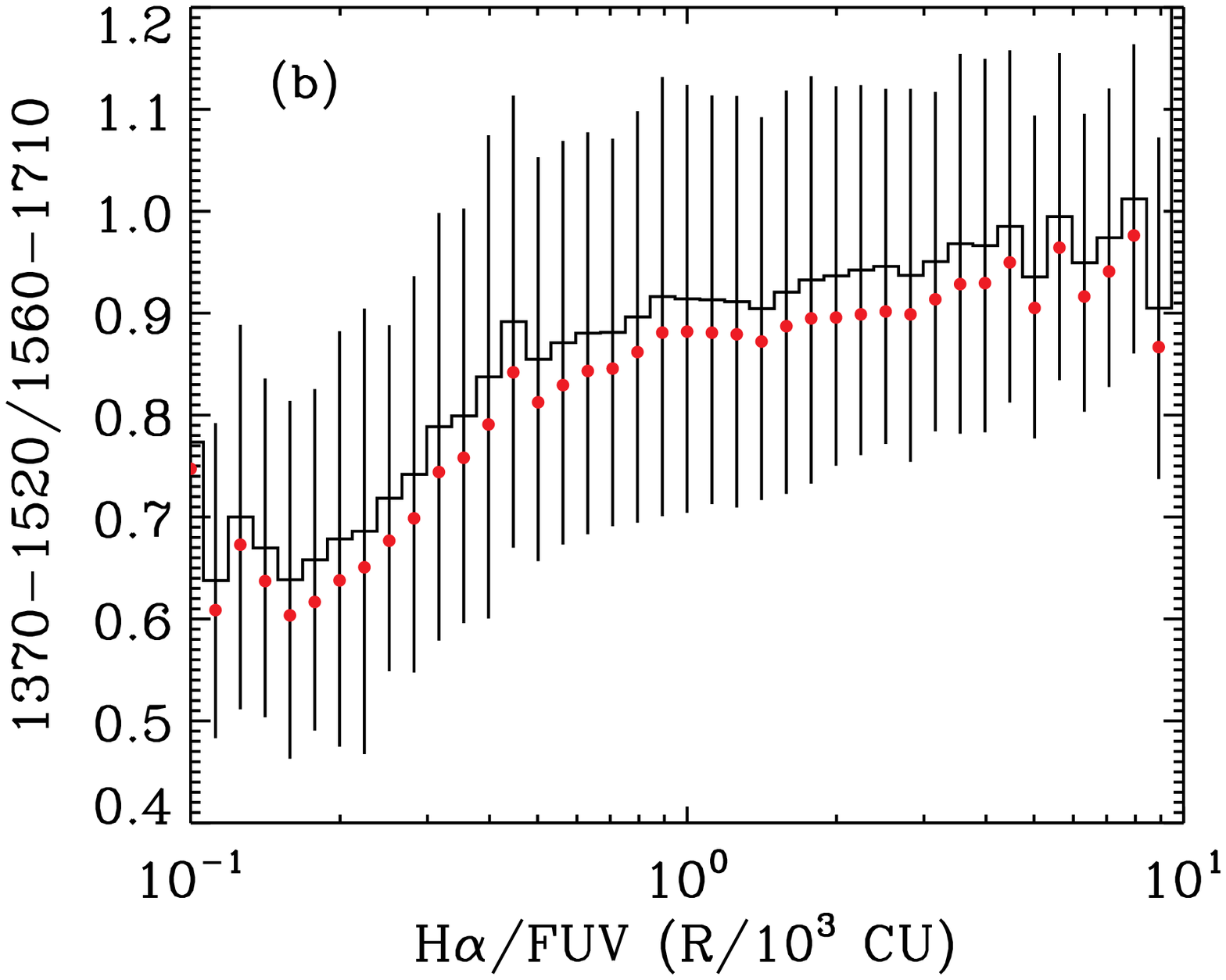}
\par\end{centering}

\caption{\label{ratio_vs_others}(a) The H$\alpha$/FUV intensity ratio versus
H$\alpha$ intensity. (b) The FUV Hardness ratio (1370--1520 to 1560--1710\AA)
versus the H$\alpha$/FUV intensity ratio. Black lines represent the
average (H$\alpha$/FUV intensity or FUV hardness) ratios for given
abscissa values after dust-extinction correction. Standard deviations
of the H$\alpha$/FUV ratios from its mean values are also shown.
Red dots represent the ratios before the correction of dust-extinction.}
\end{figure}

The mean H$\alpha$/FUV intensity ratio after and before the dust-extinction
correction as a function of H$\alpha$ intensity is shown together
with the standard deviation of ratios from the mean value for a given
H$\alpha$ intensity inverval in Figure \ref{ratio_vs_others}(a).
Uncertainty of the ratio toward an individual sightline is much smaller
than the deviation of ratios. Red and black symbols represent the
intensity ratio before and after the dust-extinction correction, respectively.
There is a general trend for the average ratio to increase with the
H$\alpha$ intensity, although it scatters a lot. Both the general
trend and large scatter are qualitatively similar to those found in
the external galaxies in \citet{Hoopes01}. The scatter in the ratio
tends to increase with the H$\alpha$ intensity. At low latitudes
and highest H$\alpha$ intensities, the H$\alpha$ radiation may be
dominated by recombination lines originated from \ion{H}{2} regions
around early O-type stars, while the FUV background is from dust-scattering
\citep{Seon2011}. Therefore, a much wider range of variability in
the H$\alpha$ intensity and the H$\alpha$/FUV ratio is found. On
the other hand, at high latitudes and low values of H$\alpha$ intensity,
much of the structure in the faint H$\alpha$ background may have
the same origin as that of the FUV background and the variation of
H$\alpha$/FUV decreases. In other words, the H$\alpha$ contribution
from dust-scattering becomes increasingly important at higher latitudes.
A similar feature was found in Figure 21 of \citet{Seon2011}, where
a linear correlation of the H$\alpha$ intensity with the FUV intensity
was found only in low intensities. In addition, at lowest H$\alpha$
intensities, only early B-type stars would contribute to both backgrounds
and the ratio would then converge to a single value. The converged
value corresponds to a spectral type of B0 -- B1.

Interestingly, line ratios {[}\ion{S}{2}{]}/H$\alpha$ and {[}\ion{N}{2}{]}/H$\alpha$,
and the H$\alpha$/FUV ratio show a similar trend with decreasing
H$\alpha$ intensity, except the line ratios increase while the H$\alpha$/FUV
ratio decreases. Mean values of the optical line ratios increase with
decreasing H$\alpha$ intensity. The ratios also show large scatter
as the H$\alpha$/FUV ratio does. The mean H$\alpha$/FUV ratio decreases
rapidly below H$\alpha$ intensity of $\sim0.7-1$ R with decreasing
H$\alpha$ intensity in Figure \ref{ratio_vs_others}(a). The mean
line ratios increase most dramatically below the same H$\alpha$ intensity
($\sim0.7-1$ R) (Figures 6, 12, and 15 in \citealp{Madsen06}). The
similar trend in both the optical line ratios and the H$\alpha$/FUV
ratio $ $may be understood as an increasing contribution of late
OB stars to the WIM as the H$\alpha$ intensity decreases.

We also plotted the hardness ratio in FUV wavelengths versus the H$\alpha$/FUV
ratio as the FUV hardness ratio might also provide an additional clue
on the spectral type of the ionizing source of the WIM. The hardness
ratio is defined as the ratio of average intensities at $\lambda\lambda1370-1520$\AA\
and $\lambda\lambda1560-1710$\AA, excluding the bright \ion{C}{4},
\ion{Si}{2}, and \ion{Al}{2} emission lines, as in \citet{Seon2011}.
We found an overall correlation between the FUV hardness ratio with
the H$\alpha$/FUV intensity ratio, as shown in Figure \ref{ratio_vs_others}(b),
where the standard deviation of the FUV hardness ratios from its mean
value for a given H$\alpha$/FUV bin is also indicated. The mean FUV
hardness ratio shows rather rapid decrease below the H$\alpha$/FUV
$\sim1$ (R/$10^{3}$ CU), corresponding to the approximate mean value
of the H$\alpha$/FUV ratio of the WIM. \citet{Seon2011} found that
a map of the hardness ratio of 1370--1520\AA\ to 1560--1710\AA\
band intensity shows that the sky is divided into roughly two parts,
which are determined by the longitudinal distribution of OB-type stars
in the Galactic plane. Therefore, the correlation between the FUV
hardness ratio and a large portion, if not all, of the H$\alpha$/FUV
ratio supports that the diffuse H$\alpha$ background is produced
by relatively late-type stars rather than \ion{H}{2} regions in the
Galactic plane.

Statistically significant correlation coefficients of 0.9 and 0.8
for Figure \ref{ratio_vs_others}(a) and (b), respectively, were obtained.
We also note that most of the H$\alpha$/FUV intensity ratios are
so concentrated to \textasciitilde{} 1 (R/10$^{3}$ CU), as can be
noted in Figure \ref{hist_ratio}, that the two-dimensional histograms
and/or contours as in Figure \ref{ha_vs_others} show strong peaks
at the H$\alpha$/FUV intensity ratio of \textasciitilde{} 1 (R/10$^{3}$
CU). Therefore, the average intensity ratio versus the abscissa in
Figure \ref{ratio_vs_others} was the best way to represent clearly
the correlations. We note that the H$\alpha$/FUV ratio map (Figure
\ref{map_ratio}) does not match perfectly with the FUV hardness ratio
map (Figure 14 of \citealt{Seon2011}). However, the maps generally
correlate well with each other, except a few regions such as near
$(l,b)\sim(30^{\circ},30^{\circ})$.

\citet{Hoopes00} and \citet{Hoopes01} compared the observed H$\alpha$/FUV
ratios with the theoretical calculations assuming that the FUV photons
are emitted from unresolved point sources in the observed regions.
However, FUV photons are well scattered by interstellar dust grains
to large distances and the diffuse FUV emission is wide-spread over
regions where no point sources are present \citep{Bowyer91,WIT1997,SCH2001,Seon2011}.
The FUV intensity used in the analysis of face-on galaxies would be
produced by in-situ point sources as well as dust-scattered light
in the regions, whereas the FUV data in the present study is dominated
by dust-scattered light originating from late O- and early B-type
stars \citep{Henry2002,Seon2011}. We also note that the spectral
type of dominant radiation source of the FUV continuum background
is consistent with that inferred from H$\alpha$/FUV intensity ratio.

It would be interesting to estimate how a large portion of the ionizing
power required for the H$\alpha$ background can be provided by late
OB stars. If the H$\alpha$ emission originates from ionized gas in
a plane-parallel geometry, the hydrogen recombination rate $r_{G}$
per cm$^{2}$ of galactic disk in the vicinity of the Sun is given
by
\[
r_{G}\approx4\times10^{6}I_{{\rm H}\alpha}(b)\times\sin|b|\ {\rm s}^{-1}\ {\rm cm}^{-2},
\]
 where $I_{{\rm H}\alpha}(b)$ is the H$\alpha$ intensity at galactic
latitude $b$ \citep{Reynolds84}. Using an approximate H$\alpha$
intensity distribution $I_{{\rm H}\alpha}(b)\approx1.0\csc|b|$ R,
which was obtained from rather limited observations, Reynolds obtained
$r_{G}\approx4\times10^{6}$ s$^{-1}$ cm$^{-2}$. However, using
the result of \citet{Hill2008} obtained from the full WIM observations,
the distribution of H$\alpha$ at high latitudes is fitted on average
by $I_{{\rm H}\alpha}(b)\approx0.625\csc|b|$ R (see also \citealt{Dong2011}).
This leads to a lower recombination rate of $r_{G}\approx2.5\times10^{6}$
s$^{-1}$ cm$^{-2}$. The Lyc luminosities per cm$^{2}$ of the Galactic
disk estimated for the solar neighborhood are $(10-29)\times10^{6}$
and $(0.5-1)\times10^{6}$ photons s$^{-1}$ cm$^{-2}$ for O and
B stars, respectively \citep{Reynolds84}. The fraction of Lyc luminosity
of O9 and O9.5 stars in total Lyc luminosity due to O stars is $\sim7-10$\%
\citep{Terzian1974,Torres-Peimbert1974}. The Lyc luminosity from
the stars later than or equal to O9 would then be $(1.2-3.9)\times10^{6}$
photons s$^{-1}$ cm$^{-2}$. Therefore, late OB stars are able to
account for at least one-half of the required ionizing power.

In fact, \citet{Elmegreen1975} considered the intensity of the H$\alpha$
background at $|b|<30^{\circ}$ and concluded that isolated \ion{H}{2}
regions surrounding early B stars could account for up to one-half
of the observed H$\alpha$ background in some regions of the Galactic
plane. He predicted that near the Galactic poles, B star \ion{H}{2}
regions would account for a mean H$\alpha$ intensity of $\sim0.1$
R. This result is consistent with our estimation.

\section{Dust Scattering}

The late OB-type stars may not have enough power to produce the whole
H$\alpha$ background if the H$\alpha$ photons originate purely from
ionized gas around late OB stars, as noted by \citet{Reynolds84}
and in Section 3. However, if dust-scattering plays an important role,
the dust-scattered H$\alpha$ photons originating from the sources
other than O stars in bright \ion{H}{2} regions would contribute
a larger portion to the H$\alpha$ background at high-latitudes and
the severe requirement of Lyc leakage from \ion{H}{2} regions in
the midplane may be potentially alleviated. In this section, we examine
how the contribution of dust-scattering could be significant in the
H$\alpha$ background.

Galactic observables may generally correlate with each other because
of the general property of a plane-parallel ISM. The correlation between
a pair of Galactic quantities would be enhanced when their sources
and radiative transfer mechanisms are analogous. If a large portion
of the diffuse H$\alpha$ intensity is due to dust-scattering, the
H$\alpha$ intensity would show a better correlation with FUV intensity
than that with other quantities, such as neutral hydrogen column density
$N$(\ion{H}{1}). On the other hand, if the radiative transfer mechanism
for the diffuse H$\alpha$ emission is not related with the dust-scattering,
the correlation between the H$\alpha$ and FUV intensities would not
be improved.

\begin{figure*}[t]
\begin{centering}
\includegraphics[clip]{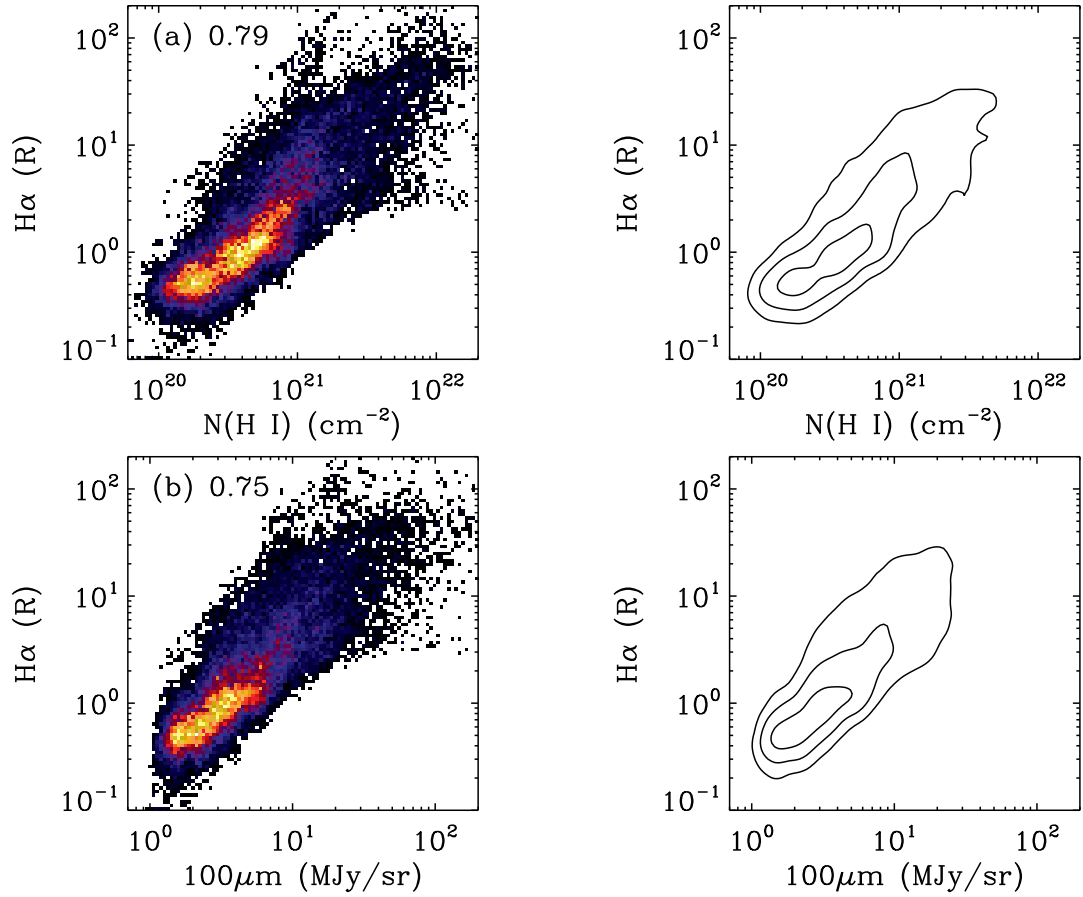}
\par\end{centering}

\caption{\label{ha_vs_others}Correlation of the diffuse H$\alpha$ background
with (a) IR 100 $\mu$m emission, and (b) neutral hydrogen column
density $N$(\ion{H}{1}). The left panels show two-dimensional histogram
and the right panels contours of the histograms. Numbers in the left
panels are the correlation coefficients estimated in logarithmic scale.
The contours correspond to 0.7, 0.3, and 0.1 of the maximum values
of the histograms. See Figure 21(c) in \citet{Seon2011} for the comparison
of the H$\alpha$ background with the FUV continuum background. The
correlation coefficient of the H$\alpha$ versus FUV intensities is
0.81 from Figure 21(c) in \citet{Seon2011}.}
\end{figure*}

Figure \ref{ha_vs_others} shows the correlation plots of H$\alpha$
intensity with 100 $\mu$m dust emission and \ion{H}{1} column density
$N$(\ion{H}{1}) in the same forms of two-dimensional histograms
and contours as in Figure 21 of \citet{Seon2011}, in which the FUV
intensity is compared with 100 $\mu$m intensity, $N$(\ion{H}{1}),
and H$\alpha$ intensity. Correlation coefficients estimated in logarithmic
scale are also shown in the figure. We find that the correlation of
H$\alpha$ intensity with FUV (shown in Figure 21(c) of \citealt{Seon2011})
is a bit better, at least qualitatively, than with the others ISM
tracers shown in Figure \ref{ha_vs_others}. The correlation relations
with 100 $\mu$m intensity and $N$(\ion{H}{1}) are a bit curved
and flaring, whereas the correlation with the FUV background is quite
straight. However, the correlation coefficient between the H$\alpha$
and FUV backgrounds is not significantly higher than the other correlation
coefficients.

\begin{figure*}[t]
\begin{centering}
\includegraphics[clip,scale=1.5]{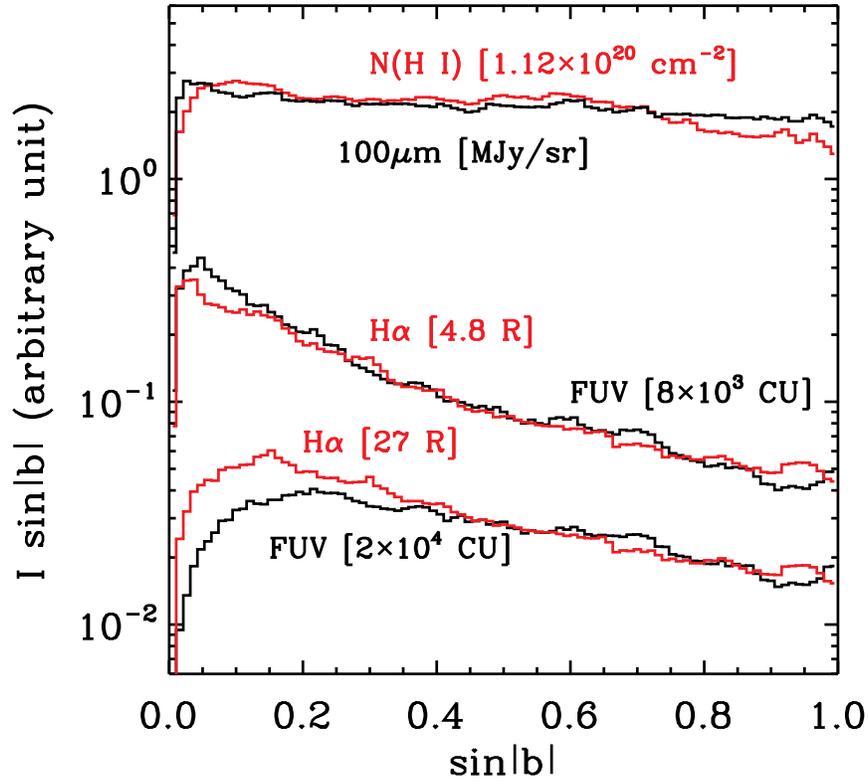}
\par\end{centering}

\caption{\label{sinb}$I\sin|b|$ versus $\sin|b|$ for the neutral hydrogen
column density $N$(\ion{H}{1}), 100 $\mu$m emission, H$\alpha$,
and FUV intensities. Curves were shifted arbitrarily for comparison
of the shapes. Top curves represent the variation of $N$(\ion{H}{1})
and the 100 $\mu$m emission in red and black colors, respectively.
Middle and bottom curves show the variation of the H$\alpha$ (red)
and FUV (black) intensities. Middle curves represent the values after
dust-extinction correction and bottom curves before the correction.
Units are denoted in square brackets.}
\end{figure*}

The fact that the correlation coefficients show no significant differences
may be caused by large scatter in the correlation relations and the
general similarity in the latitude dependence of the data sets (the
\ion{H}{1} column density, 100 $\mu$m, H$\alpha$, and FUV intensities).
Although the H$\alpha$ emissivity and \ion{H}{1} column density
have different dependences on the gas density, their latitude dependences
have the same trend with inverse-$\sin|b|$ if they arise in plane-parallel
geometries. The inverse-$\sin|b|$ dependence of the data sets dominates
the correlations in Figure \ref{ha_vs_others}. To reduce the latitude
dependence and scatter, we therefore plotted the average intensities
multiplied by $\sin|b|$ versus $\sin|b|$ within each of the $\Delta\sin|b|=0.01$
latitude intervals for $N$(\ion{H}{1}), 100 $\mu$m, H$\alpha$,
and FUV intensities in Figure \ref{sinb}. In the figure, the curves
were paired and arbitrarily moved to compare the shapes. The top two
curves compare the $N$(\ion{H}{1}) and 100 $\mu$m emission. The
middle and bottom pairs show the variations of H$\alpha$ and FUV
intensities after and before dust-extinction correction, respectively.
The $N$(\ion{H}{1}) and 100 $\mu$m emission are well described
by the plane-parallel or inverse-$\sin|b|$ law, whereas H$\alpha$
and FUV intensities deviate from the law. The deviation of H$\alpha$
and FUV intensities from the inverse-$\sin|b|$ law is mainly attributable
to the effect of dust absorption and scattering. A radiative transfer
model of diffuse Galactic light in plane-parallel dust layer shows
the similar dependence of $I\sin|b|$ on $\sin|b|$ as in the bottom
curves of Figure \ref{sinb} (\citealp{vandeHulst1969}; see also
\citealt{Wood1999} and \citealt{Seon2009b}). We also note that the
inverse-$\sin|b|$ equation of the H$\alpha$ intensity described
in Section 3 is still a good approximation because of the major dependence
of the H$\alpha$ intensity on inverse-$\sin|b|$, especially at high
latitudes, even $I\sin|b|$ is not a constant in Figure \ref{sinb}.

We note from the bottom two curves before dust-extinction correction
that the dependences of the H$\alpha$ and FUV backgrounds on the
Galactic latitude are very similar at high latitudes ($\sin|b|>0.3$)
(See also Figure 13 of \citealt{Haffner99} and Figure 9 of \citealt{Seon2011}),
whereas at low latitudes ($\sin|b|<0.3$) they show a difference.
Stronger dust-extinction in FUV would suppress the FUV intensity near
the Galactic plane, as shown in low latitudes. We therefore corrected
the dust-extinction as denoted by the middle two curves (red and black
solid lines for H$\alpha$ and FUV, respectively). The dust-extinction
correction for both backgrounds was performed using the standard approach
of \citet{Bennett2003}. The variations of both backgrounds are remarkably
similar to each other. The FUV radiation can be emitted even from
late B-type stars, whereas the H$\alpha$ emission requires excitation
from earlier stars. Therefore, the scaleheight of the FUV sources
is a bit higher than that of the H$\alpha$ sources. On the other
hands, ionized \ion{H}{2} regions must be more extended than point
sources for the FUV photons, and the dust-scattered H$\alpha$ originating
from \ion{H}{2} regions should be more extended than those from point
sources. Accordingly, the dependences of both backgrounds on the Galactic
latitude are very similar even at low latitudes, after extinction
corrections are applied.

\begin{figure*}[t]
\begin{centering}
\includegraphics[clip,scale=0.9]{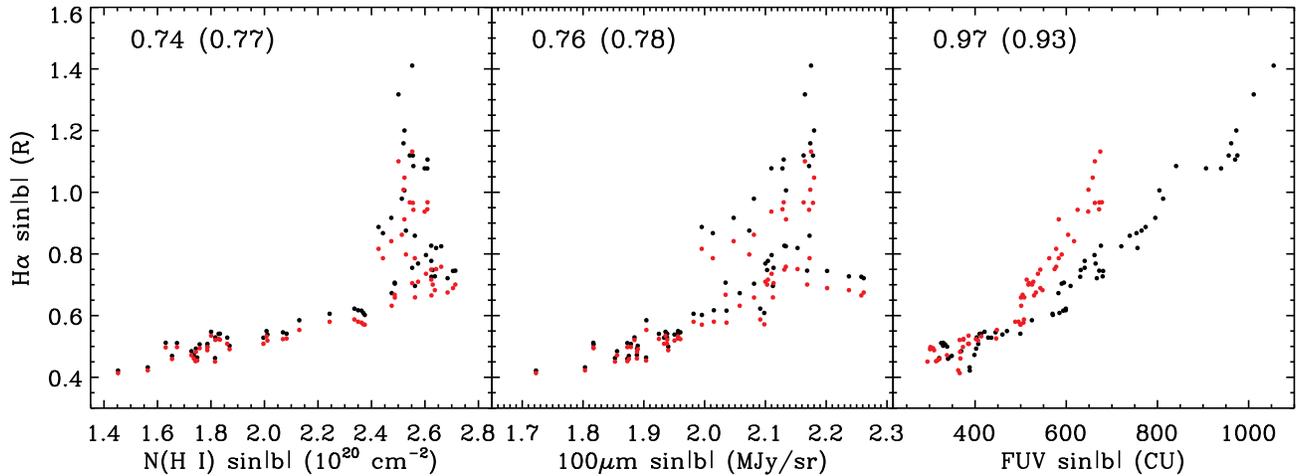}
\par\end{centering}

\caption{\label{correlation_sinb}Correlation of the diffuse H$\alpha$ intensity
with neutral hydrogen column density (left), 100 $\mu$m emission
(middle), and the diffuse FUV intensity (right). Red and black dots
represent the data points before and after the dust-extinction correction,
respectively. Numbers outside and inside the parentheses are correlation
coefficients estimated after and before the dust-extinction correction,
respectively.}
\end{figure*}

In Figure \ref{correlation_sinb}, we also examine correlations of
the diffuse H$\alpha$ intensity with $N$(\ion{H}{1}), 100 $\mu$m
emission, and the diffuse FUV background after the removal of inverse-$\sin|b|$
dependence. In the correlation plot, we used only the latitude range
of $\sin|b|>0.3$ to avoid the strong dust-extinction effect. Red
and black symbols represent the data points after and before the dust-extinction
correction, respectively. The figure also shows correlation coefficients.
It is obvious that the correlation with the FUV background is now
significantly stronger than the correlations with others. In \citet{Seon2011},
we compared the FUV continuum background with other ISM tracers and
found that the FUV intensity correlates better with the H$\alpha$
intensity than with others. The two-photon emission in the WIM is
unlikely to be the cause of the strong correlation between the H$\alpha$
and FUV backgrounds \citep{Seon2011}. We are therefore convinced
of the result that a significant portion of the diffuse H$\alpha$
emission at high latitudes is dust-scattered light originated from
elsewhere in the Galaxy.

We now estimate the fraction of dust-scattered H$\alpha$ emission.
\citet{Seon2011} derived two equations (Equations (7) and (8)) relating
the FUV intensity to 100 $\mu$m and H$\alpha$ emissions at high
latitudes. Since the constant terms in the relations are often regarded
as extragalactic background, they should be identical. The difference
between the constant terms is indeed negligible. Therefore, the relation
between the 100 $\mu$m and H$\alpha$ emission are given by $ $$I_{100\mu{\rm m}}\approx2.88\left(I_{{\rm H}\alpha}/R\right)$
MJy/sr. \citet{Witt2010} provides an empirical relation, which relates
the scattered H$\alpha$ intensity to the 100 $\mu$m emission at
high latitudes. Using these relations, we obtain an average ratio
of the scattered H$\alpha$ intensity to total intensity, $I_{{\rm H}\alpha}^{{\rm scatt}}\thickapprox0.37I_{{\rm H}\alpha}$.
Therefore, in general, about 37\% of the diffuse H$\alpha$ emission
at high latitudes would be due to dust-scattering. Contribution of
the Lyc leakage from bright \ion{H}{2} regions may be $\sim$13\%
on average if late OB stars produce about half of the H$\alpha$ emission.
However, these numbers should be referenced with caution. The scattered
portion can be much higher or lower depending on location, as shown
in \citet{Witt2010}. Since the FUV background traces the dust-scattered
light, we expect that the FUV to H$\alpha$ ratio would be more or
less proportional to the scattered fraction of H$\alpha$ background.
We note that the scattered fraction of H$\alpha$ background shown
in Figure 5 of \citet{Witt2010} is more or less anti-correlated with
the H$\alpha$/FUV ratio map of Figure \ref{map_ratio}, verifying
the expectation.

\section{Discussion}

By combing the previous results, we propose that most of the diffuse
H$\alpha$ background at high latitudes may originate from relatively
late O- and early B-stars outside the bright \ion{H}{2} regions and
from dust-scattered H$\alpha$ photons from elsewhere in the Milky
Way. The observed H$\alpha$/FUV intensity ratio in general increases
with the H$\alpha$ intensity, indicating that late OB stars could
be important contributors to the diffuse H$\alpha$. As suggested
by the strong correlation between the diffuse H$\alpha$ and FUV backgrounds,
dust-scattering may widely spread out the H$\alpha$ recombination
photons into regions where there is no ionizing source.

\subsection{Ionizing and heating sources}

The present scenario is able to explain the line ratios observed in
the WIM much more easily than or equally as well as the photoionization
models. The higher {[}\ion{S}{2}{]} $\lambda$6716/H$\alpha$ and
{[}\ion{N}{2}{]} $\lambda$6583/H$\alpha$ line ratios observed in
the diffuse H$\alpha$ regions compared to those in \ion{H}{2} regions
have been believed to be the strongest evidence against the theory
of a dust-scattering origin of the diffuse H$\alpha$ emission \citep{Reynolds1990b}.
However, we have to note that the argument has been based solely on
comparison with the line ratios from the bright \ion{H}{2} regions.
If the line ratios from the ionized regions due to the late-type OB
stars are compared with the line ratios in the WIM, the present scenario
seems to explain the trend equally as well as the previous photoionization
models do.

As already noted in Section 3, mean values of {[}\ion{S}{2}{]}/H$\alpha$
and {[}\ion{N}{2}{]}/H$\alpha$ intensity ratios in the diffuse H$\alpha$
regions are well reproduced with late OB stars \citep{Sembach2000}.
Figure 3 in \citet{Reynolds88} shows that the {[}\ion{S}{2}{]}/H$\alpha$
ratios observed in \ion{H}{2} regions around late-type stars are
higher than those found around earlier-type stars. Recently, \citet{ODell2011}
pointed out that the similarity of the line ratios between the Barnard's
Loop, the Orion-Eridanus Bubble, and the typical WIM samples is striking
and probably indicates a common set of their physical conditions.
The ionizing stars of the Barnard's Loop are O9.5 and O9 types. They
found that the photoionization models with the ionization parameter
$\log U=-3.07\sim-3.67$ and the stellar temperature $T_{{\rm star}}=31,000$
K -- 40,000 K fully enclose the space of {[}\ion{S}{2}{]}/H$\alpha$
versus {[}\ion{N}{2}{]}/H$\alpha$ occupied by Barnard's Loop, the
Orion-Eridaunus Bubble, and the WIM samples. It is also noticeable
that most of the WIM samples in their Figure 7 occupies the region
defined by a stellar temperature lower than 40,000 K, probably $\sim$
35,000 K. Another interesting fact to note is that the elemental abundances
required to explain the line ratios in the WIM are close to the B
star abundances both in \citet{Sembach2000} and in \citet{ODell2011}.
These results strongly support the late OB star origin of the diffuse
H$\alpha$ emission.

Weak {[}\ion{O}{3}{]} $\lambda$5007/H$\alpha$ and \ion{He}{1}
$\lambda$5876/H$\alpha$ emission line ratios also indicate that
the spectrum of the diffuse interstellar radiation field is significantly
softer than that from the average Galactic O star population \citep{Reynolds95,Madsen06}.

The increase of the line ratios {[}\ion{S}{2}{]}/H$\alpha$ and {[}\ion{N}{2}{]}/H$\alpha$
can be understood as a significant ioninzation contribution by late
OB stars to the WIM. However, their highest values observed in the
WIM may need other non-ionizing heating sources, such as shocks, photoelectric
heating, and/or turbulent mixing layers \citep{Reynolds1992,Slavin1993,Collins2001}.
Other heating sources may include cooling, falling galactic fountain
gas initially raised from the midplane by supernovae and microflares
from magnetic reconnection \citep{Raymond1992,Shapiro1993}. It should
be noted that the pure photoionization models by O-type stars in the
midplane also need additional heating sources to explain the highest
ratios \citep{Reynolds1999}. The emission lines due to the heating
sources that provide the highest line ratios would also be scattered
into more extended regions than those originally produced by the heating/ionizing
sources. Therefore, the dust-scattering will hamper the clear identification
of the original ionization sources.

As noted by \citet{Hoopes03}, the {[}\ion{S}{2}{]}/{[}\ion{N}{2}{]}
ratio is almost independent of ionizing stellar temperature but depends
on the ionization parameter. The photoionization models explain the
enhanced {[}\ion{S}{2}{]}/H$\alpha$ line ratios by a diluted radiation
field and thus the lowered ionization parameter. Hence the models
cannot reproduce the observed constancy of the {[}\ion{S}{2}{]}/{[}\ion{N}{2}{]}
ratios. Instead, the dust-scattering scenario can naturally explain
the constancy of the line ratios.

\subsection{Physical properties}

Dust scattering has a significant impact on line ratios which are
used to determine physical properties of \ion{H}{2} regions \citep{ODell2009,ODell2010}.
Scattered light is enhanced at shorter wavelengths, which can lead
to observed Balmer line ratios that are theoretically impossible.
This effect can also lead to overestimates of the electron temperatures
derived from the auroral and nebular line ratios of forbidden lines,
such as {[}\ion{N}{2}{]} $\lambda$5755 /{[}\ion{N}{2}{]} $\lambda$6583
which has been used to derive the temperature of the WIM. Serious
discrepancies between the H$\alpha$ intensity and radio and/or IR
intensities could be found when a large portion of the H$\alpha$
intensity is due to dust-scattering. This effect was found in the
Extended Orion Nebula (EON). \citet{ODell2009} combined 327.5 MHz
radio observations and optical spectroscopy to study conditions in
the EON. They found an increase in the ratio of emission measures
derived from the H$\beta$ line and the 327.5 MHz radio continuum
with increasing distance from the dominant photoionizing star $\theta^{1}$
Ori C, indicating the increasing contribution of a dust-scattering
component to the H$\beta$ intensity with the distance.

Recent observations using the \emph{Wilkinson Microwave Anisotropy
Probe }have found that the ratio of the free-free radio continuum
to H$\alpha$ is surprisingly low in the WIM \citep{Davies2006,Dobler2008,Dobler2009,Gold2011}.
\citet{Dong2011} proposed a three-component model consisting of a
mix of (1) hot gas currently being photoionized, (2) gas that is recombining
and cooling after removal of a photoionizing source, and (3) cold
\ion{H}{1} gas. In the standard model for explaining the observed
intensity ratios of the free-free radio continuum to H$\alpha$ and
{[}\ion{N}{2}{]} $\lambda$6583 to H$\alpha$, they assumed that
the scattered fraction of the H$\alpha$ originating from the hot
gas is 20\%. The fractions of the hot, cooling, and \ion{H}{1} gases
were found to be 22, 56, and 2\%, respectively. The model predicted
that the photoionization should switch off and the gas begins to cool
and recombine in a much shorter time ($1\times10^{5}$ yr) than O
star lifetimes ($\sim3\times10^{6}$ yr). Such a short photoionization
time scale is rather surprising in that O stars have been generally
favored as the source of ionization for the WIM. They therefore suggested
that the ionizing radiation for the WIM may be provided in large part
by runaway O and B stars, with space velocities $\gtrsim100$ km s$^{-1}$.
In the analysis, they assumed the {[}\ion{N}{2}{]}/H$\alpha$ line
ratio of $\sim0.4$ for the WIM, which is well reproduced by late
OB-type stars \citep{Sembach2000}. The ratio of free-free to H$\alpha$
is estimated to be $\psi=0.16$ kJy sr$^{-1}$ R$^{-1}$ for the WIM
with a temperature of 8000 K (Equation (11) in \citealp{Dong2011}),
which is about 1.9 times larger than the observed ratio of 0.085 kJy
sr$^{-1}$ R$^{-1}$. If we assume that $\sim$37\% of the observed
H$\alpha$ intensity is from dust-scattering and the temperature of
the WIM is 8000 K, the fractions of the hot and cooling gases are
found to be $\sim$44\% and $\sim$18\%, respectively.

The WIM is thought to have a temperature 2000 K higher than the bright
classical \ion{H}{2} regions \citep{Reynolds2001,Madsen06}. However,
\citet{ODell2011} found no evidence that the WIM components have
a systematically higher temperature than the well known higher-density
\ion{H}{2} regions. As already noted, the temperature derived from
{[}\ion{N}{2}{]} $\lambda$5755 /{[}\ion{N}{2}{]} $\lambda$6583
may be uncertain. Temperature of the Barnard's Loop is found to be
$\sim5960-6100$ K \citep{Heiles2000,ODell2011}. From the similarity
of the physical conditions between the Barnard's Loop and the typical
WIM sample, we may assume an extreme case that the temperature of
the WIM is $\sim6100$ K, and there is no contribution from either
cooling or cold \ion{H}{1} gas. The assumptions may be rather extreme,
but still plausible at some sightlines since the physical conditions
and constribution of dust scattering in individual sightlines would
generally differ from the average and vary from sightline to sightline
\citep{Madsen06,Witt2010}. The free-free to H$\alpha$ ratio of $(1-0.37)\psi(6100\ {\rm K})\sim0.085$
is then obtained, which agrees with the observed ratio. Therefore,
relatively late O and/or early B-type stars, as proposed in the present
study, appear to explain the observed values of not only optical line
ratios but also free-free to H$\alpha$ intensity ratios even without
considering a significant contribution of either cooling or cold \ion{H}{1}
gas. However, we emphasize that the aim of this study is to explain
averaged observational results and our results do not indicate that
all the diffuse H$\alpha$ emission is produced by late OB stars outside
bright \ion{H}{2} regions and dust scattering. Large scatter found
in Figure \ref{ratio_vs_others} also indicates that not only the
physical conditions and contributions of various components (dust-scattered
component, hot and cooling gases) but also the ionizing and/or heating
sources of the WIM vary significantly from sightline to sightline.

Low extinction at 2.17 $\mu$m Br$\gamma$ emission line should allow
to probe the WIM throughout the Galactic plane. A pilot survey of
the Br$\gamma$ emission carried out using the Goddard-Wisconsin near-IR
cryogenic spectrometer revealed that the volume filling fraction of
the Br$\gamma$-emitting medium is typically around 1\%, indicating
the Br$\gamma$ emission is likely related to more compact sources
\citep{Kutyrev2001,Kutyrev2004}. If the Lyc photons are indeed leaked
out of the bright \ion{H}{2} regions in the Galactic plane and photoionize
the surrounding medium to produce the diffuse H$\alpha$ emission,
as usually supposed, the filling fraction of the Br$\gamma$-emitting
gas in the Galactic plane should be equal to or higher than the filling
fraction of $\sim$10\% obtained by the H$\alpha$ observations \citep{Kulkarni1987,Reynolds1991,Peterson2002}.
This discrepancy strongly suggests that the Lyc leaked out of bright
\ion{H}{2} regions is unlikely the major source of the diffuse H$\alpha$
background and a significant portion of the H$\alpha$ background
originates from dust-scattering. More detailed observations using
hydrogen recombination lines in the near-IR wavelenths, such as Pa$\alpha$
(1.87 $\mu$m), Br$\alpha$ (4.05 $\mu$m), and Br$\gamma$, would
be needed to understand the origin of the diffuse H$\alpha$ emission.
The Multi-purpose Infra-Red Imaging System (MIRIS) is being developed
to survey the Galactic plane at Pa$\alpha$ emission line, which may
help to understand the origin of the diffuse H$\alpha$ emission \citep{Han2010}.

The volume filling fraction and scaleheight of the WIM were estimated
from the assumption that the pulsar dispersion measure and the H$\alpha$
photons probe the same ionized medium. However, as a large portion
of the diffuse H$\alpha$ background at high latitudes may be the
result of dust-scattering, the filling fraction and scaleheight must
be reexamined. We also found that some of the small-scale features
in the FUV sky coincided with the H\textgreek{a} sky, but not always.
A detailed examination of small-scale correlation between the H$\alpha$
and FUV background, which should provide better understanding of the
origin of the diffuse H$\alpha$ emission, will be presented in the
future.

\subsection{Radiative transfer mechanisms}

Assuming the conventional photoionization models of the H$\alpha$
background, the ISM may have low-density paths and voids that allow
for ionizing photons from midplane OB stars to reach and ionize gas
many kiloparsecs above the Galactic plane \citep{Wood2010}. Pathways
that provide lower than average densities to high latitudes could
enhance both scattered FUV and gas ionized by the Lyc from the same
OB stars in the Galactic plane. Stars that can reproduce most of the
FUV continuum background in the radiative transfer models \citep[e.g., ][]{WIT1997,SCH2001}
are located within only a few hundred pc from the Sun. FUV photons
are predominantly scattered in the forward direction \citep{WIT1997,SCH2001,Lee08},
implying that dust-scattering occurs mostly in between the observer
and the source. Preliminary results from our Monte-Carlo 3D radiative
transfer models for the FUV continuum background indicate that most
of the FUV continuum background originates from stars and dust located
within a volume with a size of $\pm$1 kpc centered at the Sun (Seon
et al., in preparation). The H$\alpha$ photons of the WIM are thought
to originate within 2--3 kpc distance \citep[e.g., ][]{Reynolds84},
implying that Lyc photons would also originate from a volume within
a few kpc. In this regard, we note that the dust-scattering cross-section
at 1550\AA\ is $\sim1.3\times10^{4}$ times lower than the photoionization
cross-section at 912\AA\ \citep{WD2001,Verner1995}, assuming a Galactic
dust-to-gas ratio. This suggests that FUV photons should originate
from much (probably $\sim10^{4}$ times) larger volume than H$\alpha$
photons, which is in contrast with the radiative transfer models of
the FUV continuum background. Meanwhile, the dust-scattering cross-section
for H$\alpha$ photons is lower than that for FUV (1550 \AA) by a
factor of $\sim2$ \citep{WD2001}, indicating that dust-scattering
of the FUV photons will occur at closer distances than H$\alpha$
photons. Therefore, the dust-scattering of H$\alpha$ photons provide
a much easier way to explain the diffuse H$\alpha$ background than
the conventional photoionization models.

\citet{Wood1999} found that the H$\alpha$ intensity from midplane
\ion{H}{2} regions that is scattered by dust at high latitudes is
in the range 5-20\% of the total intensity. However, we should note
that in-situ H$\alpha$ recombination photons at high latitudes would
be also scattered and this component was not included in the estimation.
In Figure 3 of \citet{Wood1999}, the scattered H$\alpha$ intensity
that is originating from the WIM through in-situ recombination is
compatible with the scattered light from bright \ion{H}{2} regions
(point sources) at some high latitudes, indicating higher fraction
of dust scattering than 5-20\%. Moreover, as noted in \citet{Witt2010},
the highly structured distribution in the ratio of H$\alpha$ scattering
to H$\alpha$ in situ emission was not accounted for in the model
of \citet{Wood1999}. The dust-scattered light of in-situ H$\alpha$
emission (originating mostly from late OB-type stars) at high latitudes,
and the complex structures of the ISM and ionizing sources may result
in a higher fraction of dust-scattering than predicted by the simplistic
calculation.

If late OB stars produce about half of the diffuse H$\alpha$ intensity,
the dust-scattered fraction of the H$\alpha$ intensity is more or
less at the same level as that due to in-situ recombination. The direct
FUV starlight and the diffuse FUV background were found to contribute
approximately the same amounts to the total FUV intensity on average
\citep{Seon2011}. The similarity of the dust-scattered fractions
between the FUV continuum and H$\alpha$ backgrounds supports again
the idea wherein their sources and radiative transfer mechanisms are
similar.

\section{Summary}

In summary, we found that the H$\alpha$/FUV intensity ratio increases
with the H$\alpha$ intensity and FUV hardness ratio, implying that
late OB-type stars mostly produce the diffuse H\textgreek{a} at high
latitudes. We also found that the H$\alpha$ correlates with the FUV
background very well. The correlations strongly suggest that a large
portion of the diffuse H$\alpha$ photons originate from late OB-type
stars and dust-scattering. We therefore propose a scenario wherein
both late OB-type stars and dust-scattering play significant roles
in the diffuse H$\alpha$ emission. The H$\alpha$ background at high
Galactic latitudes may mostly originate from late OB stars existing
outside the bright \ion{H}{2} regions and dust-scattering of the
H$\alpha$ photons. The proposed scenario appears to reproduce most
of the observed line ratios, such as average values of {[}\ion{S}{2}{]}/H$\alpha$,
{[}\ion{N}{2}{]}/H$\alpha$, {[}\ion{O}{3}{]}/H$\alpha$ and \ion{He}{1}/H$\alpha$,
and free-free to H$\alpha$ ratios. Their extreme values may need
additional heating sources, as is also needed in the pure photoionization
models. Our scenario is also in better accord with the relative constancy
of the {[}\ion{S}{2}{]}/{[}\ion{N}{2}{]} intensity ratio, which
cannot be explained with the standard photoionization models.

\acknowledgements{}

The SPEAR/FIMS is supported by NASA grant NAG5-5355 and flies on the
STSAT-1 Mission, supported of the Korea Ministry of Science and Technology.
We acknowledge the use of the Legacy Archive for Microwave Background
Data Analysis (LAMBDA). Support for LAMBDA is provided by the NASA
Office of Space Science. K.-I. S. was supported by a National Research
Foundation of Korea grant funded by the Korean government.

\clearpage{}

\end{document}